\documentclass{ws-ijmpc}

\begin{document}

\markboth{M. Ercsey-Ravasz, T. Roska, Z. N\'eda}
{Perspectives for Monte Carlo simulations on the CNN Universal Machine}


\title{Perspectives for Monte Carlo simulations on the CNN Universal Machine}

\author{M. Ercsey-Ravasz$^{1,2}$}
\address{ravasz@digitus.itk.ppke.hu}
\author{T. Roska$^{1}$}
\address{roska@itk.ppke.hu}
\author{Z. N\'eda$^{2}$}
\address{zneda@phys.ubbcluj.ro}

\address{$^{1}$ P\'azm\'any P\'eter Catholic University, Department of Information Technology,
 HU-1083 Budapest, Hungary}
\address{$^{2}$ Babe\c{s}-Bolyai University, Department of Physics, RO-400084 Cluj, Romania}

\maketitle



\begin{abstract}
Possibilities for performing stochastic simulations on the analog and fully 
parallelized  Cellular Neural Network Universal
Machine (CNN-UM) are investigated. 
By using a chaotic cellular automaton perturbed with the natural noise of the CNN-UM chip, 
a realistic binary random number generator is built. As a specific example for Monte Carlo 
type simulations, we use this random number generator and a CNN template
to study the classical site-percolation problem on the ACE16K chip. 
The study reveals that the analog and parallel architecture of the CNN-UM is very 
appropriate for stochastic simulations on lattice models. The natural trend for 
increasing the number of cells and local memories on the CNN-UM chip will definitely 
favor in the near future the CNN-UM architecture for such problems. 

\keywords{Monte Carlo simulations; Cellular Neural Networks; random number generator}
\end{abstract}

\ccode{PACS Nos.: 89.20.Ff, 05.10.Ln, 07.05.Tp}


\section{Introduction}

The Cellular Neural Network Universal Machine (CNN-UM) is an analog computer which
has on it's main processor $L \times L$ interconnected computational units (cells). The
CNN-UM which is a hardware can be easily connected to PC type computers and programmed through a special programming
language. It is believed that CNN-UM could represent in future an elegant alternative to digital computing for many
complex problems. The basic theory of cellular neural/nonlinear networks (CNN) [\refcite{1}] and the 
architecture of the CNN-UM [\refcite{2}] motivated 
hundreds of scientific papers discussing this new computational paradigm. The common 
goal of all these works was to study how CNN-UM could represent an 
alternative for the computational techniques of the future using tens of thousands of 
dynamical processors [\refcite{3,4}].   Many theoretical methods  ideal 
for the analogical (analog \&logic) architecture of the CNN-UM were developed and tested 
on different physical implementations. The already widely used applications (especially 
the ones implemented on real chips) are for image processing, robotics or sensory 
computing purposes 
[\refcite{5}]. For the success of the CNN computing within the scientific community, it is however desirable to develop, test and implement all the widely used computational methods on the emerging physical implementations. 

The standard CNN is composed by $L \times L$ analog units (cells) placed on a square lattice and interconnected 
through the $8$ nearest neighbors. Each cell is an electronic circuit characterized by an input value (voltage)$u_{i,j}$, a state value $x_{i,j}(t)$, and an output value $y_{i,j}(t)$. The 
$u_{i,j}$ input voltage can be defined for each operation, and for this operation does not change in time.
The output $y_{i,j}$ is equivalent with the $x_{i,j}$ state value in a given range. More specifically is given by a piece-wise linear function, bounded 
between $-1$ (white) and $1$ (black):
\begin{equation}
y=f(x)\equiv \frac{1}{2} ( \mid x+1 \mid - \mid x-1 \mid )
\end{equation}
 The state equation of the CNN cells results from the time-evolution of the equivalent circuit [\refcite{1}].
According to the original Chua-Yang model [\refcite{1}] (supposing nearest neighbor interactions only) 
this has the form:
\begin{eqnarray}
\frac{d x_{i,j}(t)}{d t}=-x_{i,j}( t ) + \sum_{k=-1}^{1} \sum_{l=-1}^{1} A_{k,l} y_{i+k,j+l} ( t )+  \label{state} \\
\nonumber + \sum_{k=-1}^{1} \sum_{l=-1}^{1} B_{k,l} u_{i+k,j+l}  + z_{i,j}  
\end{eqnarray}

The set of parameters $\{A,B,z\}$ are called templates representing the strength of the 
coupling between the neighbors. In standard CNN 
[\refcite{6}] $A$ and $B$ are equal for all cells, $z$ can vary in space.
There are diverse physical implementations of the CNN-UM, recently generalized as cellular wave computers 
[\refcite{7}] which could be of different types: mixed mode (analog-and-logic), emulated digital, optical, etc. 
The mixed-mode CMOS visual microprocessors [\refcite{8}]  are usually implemented according to the Full-Range model 
[\refcite{9}] which is 
a modification of this original model (the state value is also bounded and always equal with the output value). This will change however the qualitative behavior only in some very special cases which will not be considered here. 

The CNN architecture seems also promising when considering complex problems in natural sciences. Studies dealing with  partial differential equations (PDE) 
[\refcite{10,11,12,13}] or cellular automata (CA) models 
[\refcite{14,15}] prove this. Solving partial differential equations, spatially discretized on a lattice with the size of the CNN chip, is relatively easy and offers the advantage of continuity in time [\refcite{10}].  Deterministic cellular automaton 
[\refcite{14}] with simple nearest-neighbor rules are also straightforward to implement in the CNN architecture. However, many of the interesting problems in sciences deal with stochastic cellular automaton, random initial conditions or other Monte Carlo methods on lattices (spin problems, population dynamics models, lattice gas models, percolation etc...). Developing and proving the 
efficiency of stochastic simulation techniques (Monte Carlo methods) on the CNN-UM - using its stored- (or algorithmic) programmability - would be thus an important step toward its success. The aim of the present study is to investigate this possibility.  
 
It is well known that for a successful stochastic simulation the crucial starting point is a good random number generator. While computing with digital processors, the "world" is deterministic and discretized, so in principle there is no possibility to generate random events and thus really random numbers. The implemented random number generators are in fact pseudo-random number generators working with some deterministic algorithm, and it is believed that their statistics approximates well real random numbers. One can immediately realize that a first advantage of the analog architecture for Monte Carlo type simulations is that the simple deterministic and discretized "world" is lost, noise is present, and there is thus possibility for generating real random numbers. Here, we first present a realistic random number generator which uses the natural noise of the CNN-UM chip and generates random binary images with a uniform distribution of the white and black pixels. After that a method for generating binary images with any given probability of the black pixels will be described. The advantages and perspectives of these methods are discussed in comparison with digital computers. 
As an example for Monte Carlo simulations the well-known site-percolation problem is solved on the CNN-UM. 
All the methods presented in this study were tested and are properly working on an ACE16K chip.  It is found that some methods are already faster on the ACE16K than on modern PC type digital computers. Taking into account thus the natural trend that the lattice size of CNN-UM chips will be growing and that calculations on this chip are totally parallel, our results predict a promising trend.   Although, in this paper most of the spatial-temporal instructions are binary, we have to emphasize that the main strength of the Cellular Wave Computer/visual microprocessor chips are related to gray scale operators. In that area the speed advantage is in about three orders of magnitude [\refcite{13}].


\section{A more realistic binary random number generator}

There are relatively few papers presenting or using random number generators (RNG) on the CNN Universal Machine 
[\refcite{16,15,17,18}]. The already known and used ones   
are all pseudo-random number generators based on chaotic cellular automaton type update rules. In such a way these RNGs are, in reality, deterministic and for many initial conditions they might have finite repetition periods. The pseudo-randomness results only from the chaotic behavior. For Monte Carlo type simulations on digital computers we are used to pseudo RNGs and we are always aware about their limitations.
One has to mention however that the pseudo-randomness and the fact that the random number series is repeatable sometimes is also helpful. It makes easier debugging MC type programs and can be a necessary 
condition for implementing specific algorithms. 
 
In solving complicated statistical physics problems with large ensemble averages the fact that the RNG is deterministic and can have a finite repetition period limits the effectiveness of the statistics. Here, our goal is to take advantage on the fact that the CNN-UM chip is an analog device, and to use its natural noise for generating more realistic random numbers. This would assure an important advantage in Monte Carlo type simulations, relative to digital computers. 

All the pseudo RNGs developed on the CNN-UM up to the present are generating binary images with $1/2$ probability of the black and white pixels (logical $1$ and $0$ are generated with the same probability). These methods were used mainly in cryptography [\refcite{15}] and watermarking on pictures [\refcite{16}]. For many Monte Carlo 
type simulations on the CNN-UM chip we will need however to generate black and white pixels (logical $1$ and $0$) with any given probability. We will solve this task in the next section and now we will consider the simpler 
problem where the black and white pixels are generated with the same $1/2$ probability.

The natural noise of the CNN-UM chip is usually highly correlated in space and time, so it can not be used directly to obtain random binary images. Our method is based thus on a chaotic cellular automaton perturbed with the natural noise of the chip after each time step. As it will be shown later due to the used chaotic cellular automaton the correlations in the noise will not induce correlations in the generated random black and white pixels and the 
real randomness of the noise will kill the deterministic properties of the chaotic cellular automaton.     

As starting point a relatively simple but efficient chaotic CA, presented by Crounse \textit{et al.} 
[\refcite{15}] and Yalcin \textit{et al.} [\refcite{16}] called the PNP2D was chosen. This chaotic CA is based on the following update rule:
\begin{eqnarray}
x_{t+1}(i,j)=(x_t (i+1,j) \vee x_t (i,j+1) ) \oplus  x_t (i-1,j) \oplus \\
\nonumber \oplus x_t (i,j-1) \oplus x_t (i,j), 
\end{eqnarray}
where $i,j$ are the coordinates of the pixels, the index $t$ denotes the time-step, 
and $x$ is a logic value $0$ or $1$ representing white and black pixels, respectively. Symbols $\vee$ and $\oplus$ represent the logical operations or and exclusive-or (XOR), respectively. As described by the authors this chaotic CA is relatively simple and fast, it passed all important RNG tests and shows very small correlations so it is a good candidate for a pseudo-random number generator. It generates binary values $0$ and $1$ with the same $1/2$ 
probability independently of the starting condition. Our goal is to transform it into a realistic RNG.

\begin{figure}
\centerline{\epsfig{file=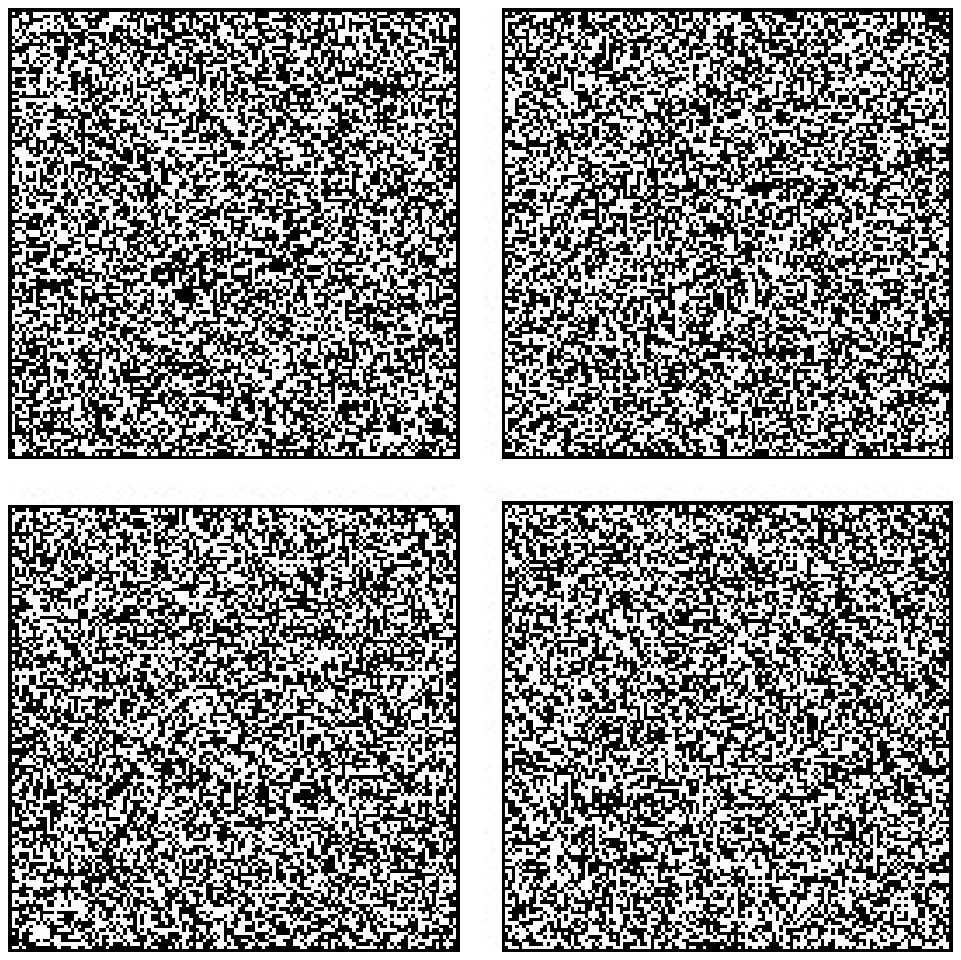,width=7.6cm}}
\caption{Four consecutive random binary images with $p=1/2$ probability of the black pixels, generated with the
presented method on an ACE16K chip.}
\end{figure}

The way we do this is relatively simple. After each time step the $P(t)$ result of the chaotic CA is perturbed with a noisy $N(t)$ binary picture (array) so that the final output is given as: 
\begin{equation}
P^{ \prime} (t)=P(t) \oplus N(t).
\end{equation}
The symbol $\oplus$ stands again for the logical operation XOR, i.e. pixels which are different on the two pictures will become black (logic value $1$). This operation assures that no matter how $N(t)$ looks like, the density of black pixels remains the same $1/2$. Because the used noisy images contain only very few black pixels (logic values $1$) we just slightly sidetrack the chaotic CA from the original deterministic path and all the good properties 
of the pseudo-random number generator will be preserved.

The $N(t)$ noisy picture is obtained by the following simple algorithm. All pixels of a gray-scale image are filled up with a constant value $a$ and a cut is realized at a threshold $a+z$, where $z$ is a relatively small value. In this
manner all pixels which have smaller value than $a+z$ will become white (logic value $0$) and the others black
(logic value $1$). Like all the logic operations this operation can be also easily realized on the CNN-UM. 
Due to the fact that the used CNN-UM chip is an analog device, there always will be natural noise on the gray-scale image.  Choosing thus a proper $z$ value one can generate a random binary picture with few black pixels. Since the noise is time dependent and generally correlated in time and space, the $N(t)$ pictures might be strongly correlated but will fluctuate in time. These time-like fluctuations can not be controlled, these are caused by real stochastic processes in the transistor circuits of the chip and are the source of a convenient random perturbation for our RNG based on a chaotic CA.  

\begin{figure}
\centerline{\epsfig{file=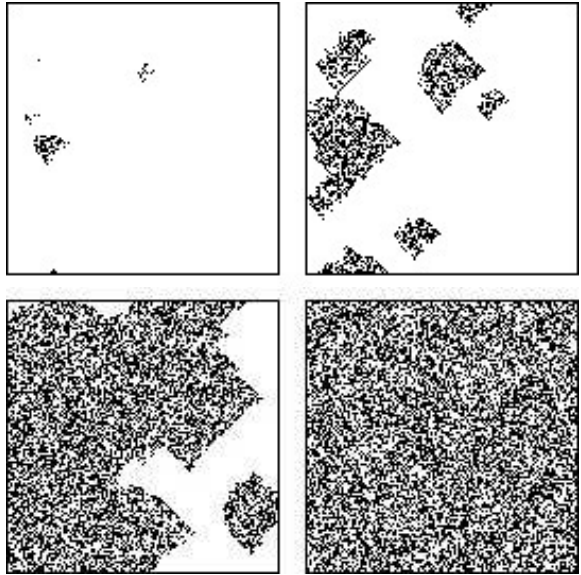,width=7.6cm}}
\caption{Illustration of the non-deterministic nature of the generator. The figure presents the $P^{\prime}_1(t) \oplus P^{\prime}_2(t)$ picture resulting from two different realizations with the same 
initial condition $P_1(0)=P_2(0)$, in the $t=10, 25, 50, 75$ iteration steps, respectively.}
\end{figure}

We performed our experiments on the ACE16K chip (with $128 \times 128$ cells) included in a Bi-i v2 
[\refcite{19}]. We have chosen
the values $a=0.95$ and $z=-0.012$. On this chip we observed that the noise is bigger when $a$ is close to $1$, and that was the reason why we have chosen $a=0.95$. The motivation for the negative $z$ value is the following. Our experiments revealed a relatively strong negative noise on gray-scale images. Due to this negative noise, once a gray-scale picture with a constant $a$ value ($0<a<1$) is generated, the measured average pixel value on the picture will always be smaller than $a$. The chosen small negative
value of $z$ ensured getting an $N(t)$ array with relatively few black pixels. In case the noise on the gray-scale picture is different (a different chip for example) one will always find a proper value for $z$.   
Choosing the above $a$ and $z$ values, it was tested that no correlations are observed in the generated patterns and
the density of black and white pixels are the same. Some consecutive random images generated by this method 
with the above mentioned $a$ and $z$ parameters are shown in Fig. 1.  
Perturbing the CA with this noise assures also that starting from the same initial state our realistic RNG after the same time-steps will yield different results $P^{\prime}_1(t)$, $P^{\prime}_2(t)$, $P^{\prime}_3(t)$ etc... In Fig. 2 starting from the same initial condition (initial random binary picture) we compare for several time steps the generated patterns. In this figure we plot the image resulting from an XOR operation performed on the $P^{\prime}_1(t)$ and $P^{\prime}_2(t)$ pictures.   In case of a simple deterministic CA this operation would yield a completely white image for any time step $t$. As visible from Fig. 2 in our case almost the whole picture is white in the beginning showing that the two results are almost identical, but as time passes the small $N(t)$ perturbation propagates over the whole array and generates completely different binary patterns. For $t>70$ time-steps the two results are already totally different. The CNN codes written for the generation of random binary patterns can be downloaded from the home-page dedicated to this study 
[\refcite{20}].

During our experiments it was also observed that in some cases the chip already introduces noise without the 
imposed $N(t)$ perturbation. On the used Bi-i v2 configuration we have only $2$ local logic memories. For avoiding the relatively slow data transfer between the ACE16K chip and the incorporated DSP we have to save some of the binary images in local analog memories and through this transformation process, rarely but some pixels might change.  This effect introduces similar perturbations like the one described in our method. This random perturbation would not be observable of course, if the number of local logic memories would be bigger and no data transfer to local 
analog memories would be necessary.

\begin{figure}
\centerline{\epsfig{file=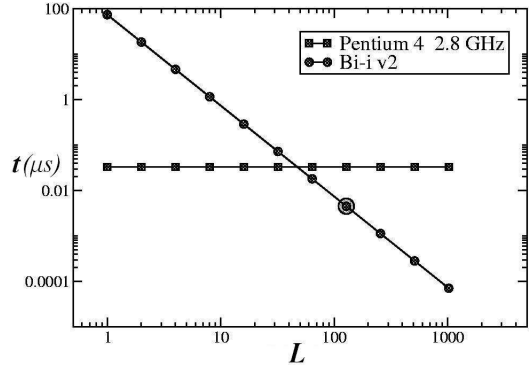,width=7cm}}
\caption{Time for generating one single binary random value on a Pentium 4 PC with $2.8GHz$ and on the
used CNN-UM chip, as a function of the CNN-UM chips size which may change in the future. Results on the actual ACE16K chip with L=128 is pointed out with a bigger circle. The results for $L>128$ chip sizes are extrapolations.}
\end{figure}

A key issue that will decide the future of the CNN-UM chips for Monte Carlo type simulations is the speed of the implemented RNG. We compared thus the speed of the above presented RNG with RNGs under C++ on normal digital computers working on a RedHat 9.0 LINUX operating system. In our experiments the time for generating a new and roughly independent random binary image on the ACE16K chip is $116 \mu s$. This means that for one single random binary value we need $116/L^2 \mu s$, where $L$ is the lattice size of the chip. In our case $L=128$, so the time needed for one random binary value is roughly $7 ns$. On a Pentium 4, $2.8 GHz$ machine this time is approximately $33 ns$. We can see thus that parallel processing makes CNN-UM already faster, and considering the natural trend that lattice size of the chip will grow, this advantage will amplify in the future. The estimated computation time for one random binary value as a function of chip size and in comparison with a Pentium 4, $2.8 GHz$ PC computer is plotted in Fig.3. 

Including more local memories on the chip will also increase the speed of the algorithms. On the actual version of the CNN-UM we have only $2$ local memories for binary images and $8$ for gray-scale images, and as we mentioned earlier 
with this configuration our algorithm needs to transfer binary pictures to gray-scale ones.

For the special cases when one needs repeatable series of random numbers the simple chaotic cellular automaton
without perturbation is sufficient. This will work in a repeatable manner only if the presently observed mistakes on the chip will be eliminated.


\section{Generating binary values with any given probability}

Up to now we considered that black and white pixels ($1$ and $0$) must be generated with equal $1/2$ probabilities. For the majority of the Monte Carlo methods this however is not enough and one needs to generate binary values with any probability $p$. On digital computers this is done by generating a real value in the interval $[0,1]$ with a uniform distribution and making a cut at $p$. Theoretically it is possible to implement similar methods on CNN-UM by generating a random gray-scale image and making a cut-off at a given value. However, on the actual chip it is extremely hard to achieve a gray-scale image with a uniform distribution of the pixel values between $0$ and $1$ (or $-1$ and $1$). 
Our solution for generating a random binary image with $p$ probability of the black pixels is by using many independent binary images with $p=1/2$ probability of the black pixels. We reduce thus this problem, to the
problem already solved in the previous section.

Let $p$ be a number between $0$ and $1$,
\begin{equation}
p=\sum_{i=1}^8  x_i\cdot 1/{2^i}
\label{p}
\end{equation}
represented here on $8$ bits by the $x_i$ binary values. One can approximate a random binary image with any 
fixed $p$ probability of the black pixels by using $8$ images $I_i$, with probabilities $p_i=1/{2^i}$, $i\in\{1$,\dots,$8\}$ of the black pixels and satisfying the condition that $I_i \cap I_j=\emptyset$ (relative to the 
black pixels) for any $i\neq j\in \{1$,\dots,$8\}$. Once these $8$ images are generated one just have to unify (perform OR operation) all $I_i$ images for which $x_i=1$ in the expression (\ref{p}) of $p$.

Getting these $8$ basic $I_i$ images is easy once we have $8$ independent images ($P_i$) with $p=1/2$ probabilities of the black pixels. Naturally $I_1=P_1$. The second image with $1/4$ probability of the black pixels 
is generated as: $I_2=\overline{I_1}\cap P_2$, where $\overline{I_i}$ denotes the negative of image $I_i$ ($\overline{I}=$ NOT $I$). In this manner the probability of black pixels on this second image $I_2$ will be $p_2=p_1\cdot p_1=1/4$ and condition $I_1\cap I_2=\emptyset$ is also satisfied. Adding now the two images $I_1$ and $I_2$ we obtain an image with $3/4$ density of black pixels $I^{\prime}_2=I_1\cup I_2$. This $I^{\prime}_2$ image is used than to construct $I_3$: $I_{3}=\overline{I^{\prime}_2}\cap P_{3}$. It is immediate to realize that $I_3$ has a density 
$1/8$ of black pixels and that $I_3 \cap I_2=I_3 \cap I_1=\emptyset$. In the next step in a similar manner we construct 
$I^{\prime}_3=I_1 \cup I_2 \cup I_3$ and $I_4=\overline{I^{\prime}_3} \cap P_4$. The method is repeated recursively until all $I_i$ are obtained. 

The above algorithm implemented on the ACE16K chip reproduced the expected probabilities nicely. The differences between the average density of black pixels (measured on $1000$ images) and the expected $p$ probability were between $0.01\%$ and $0.4 \%$. Normalized correlations in space between the first neighbors were measured between $0.05 \%$ and $0.4 \%$, correlations in time between $0.7 \%$ and $0.8 \%$.

\begin{figure}
\centerline{\epsfig{file=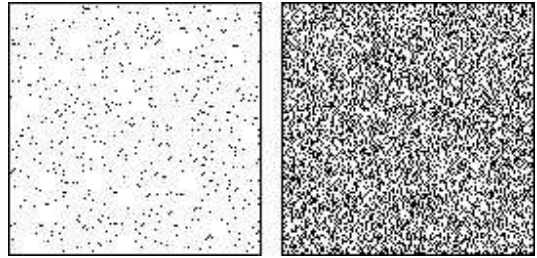,width=7cm}}
\caption{Random binary images with $p=0.03125$ (left) and $p=0.375$ (right) probability of black pixels, both of them obtained on the ACE16K chip.}
\end{figure}

Two random images with different probabilities of black pixels ($p=1/{2^5}=0.03125$ and $p=1/{2^2}+1/{2^3}=0.375$) 
are shown in Fig.4. Since the presented method is based on our previous realistic RNG the images and binary random numbers generated here are also non-deterministic. The computer codes for this algorithm can be also
downloaded from the home-page dedicated to this study 
[\refcite{20}]. 

The speed of the algorithm depends in a great measure on the probability $p$. For example, if the biggest index for which $x_i=1$ is only $3$, we need only $3$ independent random images ($P_i$) and also the recursive part of the algorithm is shorter. In the case when we need $8$ random images, the algorithm is at least $8$ times slower than for the $p=1/2$ case. However, in general we rarely need $8$ images. Working on $24$ bits
will of course further increase the calculation time. In the presented algorithm one would need
up to $24$ operations to generate the targeted random number.

It worth mentioning also, that the possible values of $p$ can be varied in a more continuous (smooth) manner, if 
$p$ is represented not on $8$ but on arbitrary $n$ bits. In this manner one has to generate $n$ binary images and
the computations on these pictures will become also more time-costly. However, again the 
increasing trend for the chip size and number of local memories on the chip will offer even in this case a big advantage in favor of the CNN-UM chips in the near future. 

    
\section{Solving a classical problem: the site percolation}

The realistic binary random number (image) generator developed in the previous sections can be used with success for 
Monte Carlo type simulations on the CNN-UM chip. As an example in this sense, here we study the 
classical site-percolation problem on the CNN-UM.

Percolation type problems are very common in many areas of sciences like physics, biology, sociology and chemistry
(for a review see e.g. [\refcite{21}]). Different variants of the problem (site percolation, bond percolation, directed percolation, continuum percolation etc.) are used for modeling various natural phenomena
[\refcite{22}]. As an example, the well-known site percolation problem is widely used for studying the conductivity or mechanical properties of composite materials, the magnetization of dilute magnets at low temperatures, fluid passing through porous materials, forest fires or
propagation of diseases in plantations etc. The site-percolation model exhibits a second order geometrical 
phase transition and it is important also as a model system for studying critical behavior [\refcite{23}].  
The site-percolation problem can be formulated as follows: we activate the sites of a lattice with a fixed $p$ probability and than we detect whether there is a continuous path on activated sites through the neighbors from one side of the lattice to the opposite one. In most of the cases the neighbors for the detection of percolation are considered to be nearest neighbors, but one can consider the problem also for both nearest and next-nearest or even higher order neighbors.  The percolation problem can be formulated in an analog manner on random binary images. 
After generating a random binary image with $0 \le p \le 1$ density of the black pixels, one is interested whether it is possible or not to go from one side of the picture to the opposite side through activated and neighboring pixels. If there is a path that satisfies this condition, it is considered that the black (activated) pixels percolate through the lattice. For the same $p$ density of black pixels it is obvious that for some random realization there will be percolation and for others there is not. For a mathematical study of the percolation problem one can define and study thus the $\rho$ probability of having a percolation, which is obtained from studying many random realizations with the same $p$ black (activated) pixels density 
[\refcite{24}]. The $\rho$ probability that a random image percolates depends of course on the $p$ density of black pixels. For an infinite lattice (image size) there is a critical density $p_c$ under which the probability of percolation goes to zero ($ \rho \rightarrow 0$), and above $p_c$ it has a non-zero probability (for 
$p \rightarrow 1$ we get of course always $\rho \rightarrow 1$). In the thermodynamic limit (infinite system size) we have thus a clear geometrical phase transition-like phenomena. For a finite system the abrupt change for $\rho (p)$ in the vicinity of $p_c$ is much smoother. 

There are several important 
quantities that are in general studied for this phase-transition. The main quantities under investigation 
are some critical exponents, the shape of $\rho (p)$ and the value of $p_c$. In the simple nearest-neighbor case 
and for the two-dimensional case the site-percolation problem is analytically solvable. For more complicated cases
and for higher dimensionality there are many available analytic approximation methods, but the most commonly 
used study method is Monte Carlo simulation. Here we will show how it is possible to determine the shape of the 
$\rho (p)$ curve on a fixed-size lattice by computer simulations on the CNN-UM. We study the simple site-percolation
problem on binary images where for percolation we consider both the nearest-neighbors and the next
nearest-neighbor pixels. From the architecture of the ACE16K CNN-UM chip it results that the used lattice is a square one, and from the considered neighbor definition results that each pixel has $8$ other neighbors.     

\begin{figure}
\centerline{\epsfig{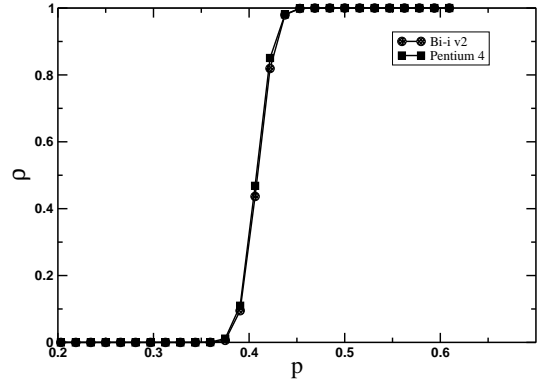}}
\caption{Simulated site-percolation probability as a function of the density of black pixels. Circles are 
results obtained on the CNN-UM chip, squares are simulation results on a normal PC type digital computer}
\end{figure}

For determining the $p_c$ critical density and the $\rho(p)$ curve one has to find the percolation probability for many different densities of the black pixels. On the CNN-UM for a random binary image the defined percolation can be detected using a  template (the parameters in eq. \ref{state}), called {\em figure reconstruction} with parameters: 
\begin{displaymath}
\mathbf{A} = 
\left( \begin{array}{ccc}
 0.5 & 0.5 & 0.5 \\ 
 0.5 & 4 & 0.5 \\
 0.5 & 0.5 & 0.5 
\end{array} \right) ,
\mathbf{B} = 
\left( \begin{array}{ccc} 
0 & 0 & 0 \\ 
0 & 4 & 0 \\ 
0 & 0 & 0 
\end{array} \right) ,  z=3.  
\end{displaymath}
The input picture of the template will be the actual random image and the initial state will contain only the first row of the image. Analyzing the template one can easily realize that pixels which have an input value equal to $1$ (are black) and have at least one neighbor with state value $1$ will become black. In this manner a flow starts from the first row making black all pixels which were black on the input picture  and are connected through (the $8$ nearest) neighbors to the first row. If on the output will remain black pixels in the last row, then percolation exists. This simple template is a function in the image processing library of the Bi-i v2 [\refcite{25}].  Applying thus this template on many random images 
generated through the methods presented in the previous section, it is possible to study the classical site-percolation problem. The computer code written for this application can be downloaded again from [\refcite{20}].

Results for the $\rho(p)$ curve obtained on the ACE16K chip is plotted with circles in Fig.5. On the same graph it is also sketched with square symbols the MC simulation results obtained on a digital Pentium 4, $2.8GHz$ computer, using a recursion-type algorithm for the detection of percolation. The lattice size in both cases is $128 \times 128$. The results in both cases are averaged for $10000$ different random images per each density value. The two curves show a very good agreement. The density values on which the simulations were done 
are $p_i=i/{2^6}$, because for the generation of random pixels $p$ has being taken as a $6$ bit value number.  The percolation threshold resulting from
the simulated $\rho(p)$ curves  are in good agreement with the accepted $p_c$ critical value, 
which for this case (site-percolation on a square lattice with $8$ neighbors) is $p_c=0.407$. 

\begin{figure}
\centerline{\epsfig{file=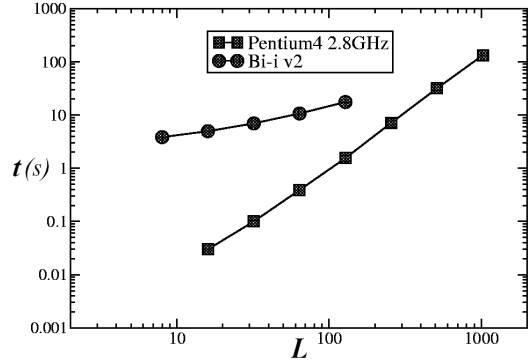,width=7cm}}
\caption{Time needed for detecting percolation on 1000 images as a function of the image linear size. Circles are results
obtained on an ACE16k CNN-UM and squares are simulation times on a Pentium 4 $2.8GHz$ PC.}
\end{figure}

Regarding the speed of these Monte Carlo type simulations performed on digital computers and on the ACE16K chip the following facts are observable: (i) with the actual chip size  (ACE16K with $L=128$) CNN-UM is still slower than a digital computer with $2.8 GHz $, (ii) on CNN-UM the time needed for detecting percolation grows linearly with the linear size of the respective image, while on digital computers it scales with the square of the lattice linear size. Increasing thus the size of the chip will definitely favor the CNN-UM architecture for such Monte Carlo type simulations on lattices. 
This trend results clearly from Fig. 6, where on a log-log scale the simulation times are compared 
on digital and CNN computers for different lattice (chip) sizes.
We have to mention also that some of the CNN instructions are not efficient on the present
ACE 16k chip.  The new chips, under fabrication are capable of performing them much more efficiently
increasing thus further the computational speed.


\section{Discussions and outlooks}

We have studied thus the possibility of performing Monte Carlo type simulations on CNN-UM chips. As a starting
tool for this task a realistic binary random number generator was built. This random number generator is based on a chaotic CA which is perturbed with the natural noise of the chip and it is able to generate realistic random binary images with $1/2$ probability of the black pixels. Starting from this random number generator it has been shown
how it is possible to generate random binary images with any desired $p$ probability of the black pixels. The great advantage of the introduced binary random number generator is the used non-deterministic algorithm and the completely parallelized manner in which the binary random number generation is made on 
all the $L \times L$ cells of the chip. It has been argued that 
this parallel architecture of the CNN-UM chip assures a big advantage and perspectives of the CNN computers relative 
to the digital ones. Some algorithms needed for stochastic simulations (like binary random number generation) are
already faster on the CNN-UM chip than on digital computers. The presently observed trend of increasing the number of cells, dimensionality and local memories on the CNN-UM chip, will definitely favor in the future the CNN-UM architecture for Monte Carlo simulations.       

As a specific example for Monte Carlo simulations on the CNN-UM chip we simulated the well-known site-percolation problem. For this application it was found however that simulations on the presently available
digital computers are still faster than the one performed on the CNN-UM chip. The reason for this is that the
percolation detection on the CNN-UM is realized through an algorithm which increases linearly with system sizes ($L$).   
The strong advantage of the CNN-UM is presently evident only for those problems where the algorithm runs totally 
parallel on all the cells of the chip. Our binary random number generator is a good example in this sense.  
Another example from statistical physics that seems to favor already the CNN-UM architecture is the Metropolis or Glauber 
Monte Carlo method on Ising type systems. Implementing these simulations on the CNN-UM chip would be a next step in the area of stochastic simulations on the CNN-UM.

\section*{Acknowledgments}
The support of the Jedlik Laboratories of the P. P\'azm\'any Catholic 
University is gratefully acknowledged.

\end{document}